\documentclass{sig-alternate}
\usepackage{amssymb,amsmath}

\newcommand{\I}{\mathbb{I}}

\renewcommand{\vec}[1]{\mathbf{#1}}
\newcommand{\Sim}{\mathrm{Sim}}

\newcommand{\ExpNCall}[1]{\text{Exp-}#1\text{-Call@}k}
\newcommand{\TlessK}{T_{\!\!\:k\text{-}1}}
\def\argmax{\operatornamewithlimits{\!arg \;\! max\,}}

\begin{document}

\conferenceinfo{SIGIR'12,} {August 12--16, 2012, Portland, Oregon, USA.} 
\CopyrightYear{2012} 
\crdata{978-1-4503-1472-5/12/08} 
\clubpenalty=10000 
\widowpenalty = 10000

\title{On the Mathematical Relationship between Expected n-call@k and 
the Relevance vs. Diversity Trade-off}

\numberofauthors{3}

\author{
\alignauthor
Kar Wai Lim\\
\affaddr{ANU \& NICTA}\\
\affaddr{Canberra, Australia}\\
\email{karwai.lim@anu.edu.au}
\alignauthor
Scott Sanner\\
\affaddr{NICTA \& ANU}\\
\affaddr{Canberra, Australia}\\
\email{ssanner@nicta.com.au}
\alignauthor
Shengbo Guo\\
\affaddr{\mbox{Xerox Research Centre Europe}}\\
\affaddr{Grenoble, France}\\
\email{shengbo.guo@xrce.xerox.com}
}

% ===============================================================================

\maketitle
%\begin{abstract}
%
%\end{abstract}
%% A category with the (minimum) three required fields
%\category{H.3.3}{Information Search and Retrieval}{Retrieval Models}
%\terms{Algorithms}
%\keywords{diversity, set-level relevance, maximal marginal relevance}

% ===============================================================================

\begin{abstract}
It has been previously noted that optimization of the $n$-call@$k$
relevance objective (i.e., a set-based objective that is 1 if at least
$n$ documents in a set of $k$ are relevant, otherwise 0) encourages
more result set diversification for smaller $n$, but this statement
has never been formally quantified.  In this work, we explicitly
derive the mathematical relationship between \emph{expected
$n$-call@$k$} and the \emph{relevance vs. diversity trade-off} ---
through fortuitous cancellations in the resulting combinatorial
optimization, we show the trade-off is a simple and intuitive
function of $n$ (notably independent of the result set size $k \geq n$), 
where diversification increases as $n \to 1$.
\end{abstract}

%\terms{Algorithms}

\section{Relevance vs. Diversity}

\emph{Subtopic retrieval} --- ``the task of finding documents that
cover as many \emph{different} subtopics of a general topic as
possible''~\cite{zhai03Beyond} --- is a
motivating case for diverse retrieval.  
%
%That is, if a query has
%multiple facets that should be covered by a result set, or a query has
%multiple ambiguous interpretations, then a retrieval algorithm should
%try to ``cover'' all of these subtopics in its result set.
%
%One of the basic tenets of set-based information retrieval is to
%minimize redundancy, hence maximize diversity, in the result set to
%increase the chance that the results will contain items relevant to
%the user's query~\cite{goffman64}.  Hence, \emph{diverse retrieval}
%can be defined as a \emph{set-level} retrieval objective that takes
%into account inter-document relevance dependences when producing a
%result set relevant to a query.
%
One of the most popular result set diversification methods is Maximal
Marginal Relevance (MMR)~\cite{carbonell98MMR}.  
Formally, given an
\emph{item set} $D$ (e.g., a set of documents) where retrieved items
are denoted as $s_i \in D$, we aim to select an optimal subset of
items $S_k^* \subset D$ (where $|S_k^*| = k$ and $k < |D|$)
\emph{relevant} to a given query $\vec{q}$ (e.g., query terms) with
some level of \emph{diversity} among the items in $S_k^*$.  MMR
builds $S_k^*$ in a greedy manner by choosing the next optimal
selection $s_k^*$ given the set of $k-1$ optimal selections
$S_{k-1}^* = \{ s_1^*, \ldots, s_{k-1}^* \}$ (recursively defining
$S_k^* = S_{k-1}^* \cup \{ s_k^* \}$ with $S_0^* = \emptyset$)
as follows:
%MMR chooses $s_k^*$
%greedily according to the following criteria:
\begin{equation}\label{eq:MMR}
 s_k^* = \hspace{-.3mm} \argmax_{s_k \in D \setminus S_{k-1}^*} [ \lambda(\Sim_{1}(\vec{q},s_k))\hspace{-.3mm}-\hspace{-.3mm}(1-\lambda)\max_{s_i \in S_{k-1}^*} \Sim_{2}(s_i,s_k) ].
\end{equation}
Here, 
$\lambda \in [0, 1]$, metric $\Sim_{1}$ measures
query-item relevance, and metric $\Sim_{2}$ measures the similarity
between two items.
%In the case of $s_1^*$, the $\max$ term is omitted.

%In MMR, we note that the $\lambda$ term explicitly controls the
%trade-off between relevance and diversity.

%In MMR, $\lambda$
%has been traditionally set in an ad-hoc manner or
%in recent work, learned in a query-specific
%way from data~\cite{santos2010selectively}.  

Presently, little is formally known about how a particular selection
of $\lambda$ relates to the overall \emph{set-based relevance objective}
being optimized.  However, it has been previously noted that the
$n$-call@$k$ set-based relevance metric (which is 1 if at least $n$
documents in a set of $k$ are relevant, otherwise 0) encourages
diversity as $n \to 1$~\cite{chen06Less,wang09PortfolioTheory}.
Indeed, Sanner \emph{et al.} \cite{sanner11}~have shown that optimizing
\emph{expected $n$-call@$k$} for $n=1$ corresponds to 
$\lambda = 0.5$ --- we extend
this derivation to show that $\lambda = \frac{n}{n+1}$ 
for arbitrary $n \geq 1$ 
(independent of result set size $k \geq
n$).  This result precisely formalizes a relationship
between $n$-call@$k$ and the relevance vs. diversity
trade-off.

\section{Relevance Model and Objective}

%%%%%%%%%%%%%%%%%%%%%%%%%%%%%%%%%%%%%%%%%%%%%%%%%%%%%%%%%%%%%%%%%%
\begin{figure}[t!]
\centerline{\includegraphics[scale = .56]{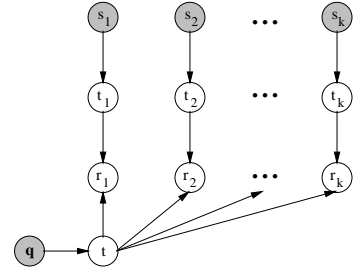}}
\vspace{-2mm}
\caption{Latent subtopic binary relevance model.}
\vspace{-3mm}
\label{fig:gm}
\end{figure}
%%%%%%%%%%%%%%%%%%%%%%%%%%%%%%%%%%%%%%%%%%%%%%%%%%%%%%%%%%%%%%%%%%

We review the \emph{probabilistic subtopic model of binary relevance}
from~\cite{sanner11} shown as a directed graphical model 
in Figure~\ref{fig:gm}.
Shaded nodes represent observed variables, unshaded nodes are
latent.  Observed variables are the query terms
$\vec{q}$ and selected items $s_i$ (where for $1 \leq i \leq k$, $s_i
\in D$).  For the subtopic variables, let $T$ be a discrete subtopic
set.  Then $t_i \in T$ represent subtopics for respective
$s_i$ and $t \in T$ represents a subtopic for query $\vec{q}$.  The
$r_i$ are $\{ 0,1\}$ variables that indicate if respective selected
items $s_i$ are relevant ($r_i=1$).

The conditional probability tables (CPTs) are as follows: $P(t_i|s_i)$
and $P(t|\vec{q})$ respectively represent the subtopic distribution
for item $s_i$ and query $\vec{q}$.  %The remaining CPTs are 
For the $r_i$ CPTs, using $\I[\cdot]$ as a $\{0,1\}$ indicator 
function (1 if $\cdot$ is true), 
item $s_i$ is deemed \emph{relevant}
\emph{iff} \emph{its subtopic $t_i$ matches query subtopic $t$}:

\vspace{-2.5mm}
{\footnotesize
\begin{align*}
P(r_{i}=1|t, t_{i}) & \; = \; \I[t_{i} = t]
\end{align*}}
%Here, $\I[\cdot]$ is $1$ when its argument is true and $0$ otherwise.
We next define $R_k = \sum_{i=1}^k r_i$, where $R_k$ is 
the number of relevant items from the first $k$ selections.  
Reading $R_k \geq n$ as $\I[R_k \geq n]$, 
we express the \emph{expected $n$-call@$k$} objective as 
\begin{align*}
  \ExpNCall{n}(S_k,\vec{q})
  = \mathbb{E}[R_k\geq n|s_1,\dots,s_k,\vec{q}] .
\end{align*}
%Since jointly optimizing $\ExpNCall{n}(S_k,\vec{q})$ is NP-hard, we
%take a greedy approach similar to MMR where we choose the best $s_k^*$
%assuming that $S_{k-1}^*$ is given.

\section{Main Derivation and Result}

%We cast the optimization of $\ExpNCall{n}(S_k,\vec{q})$
%in a form similar to MMR in~\eqref{eq:MMR} to 
%determine the correspondence between $\lambda$ and the
%result of this derivation. 
Taking MMR's greedy approach, we select $s_k$ given 
$S_{k-1}^*$:\footnote{We present a derivation summary; 
A full derivation may be found in an online 
appendix at the authors' web pages.}
%assuming that $S_{k-1}^*$ is already chosen:
\begin{align}
  s_k^* & = \argmax_{s_k} \mathbb{E}[R_k\geq n|S_{k-1}^*,s_k,\vec{q}] \nonumber\\[-1mm]
  & = \argmax_{s_k} P(R_k\geq n|S_{k-1}^*,s_k,\vec{q}) \nonumber 
\end{align}
%In the second step we have 
%exploited the $\{0,1\}$ nature of $R_k \geq n$ to rewrite $\ExpNCall{n}$
%directly as a probabilistic query.
This query can be evaluated w.r.t.\ our latent subtopic binary relevance
model in Figure~\ref{fig:gm} as follows, where we marginalize out
all non-query, non-evidence variables $T_k$ and define
$T_k\!=\!\{t,t_1,\dots,t_k\}$ and 
$\sum_{T_k} \circ = \sum_t \sum_{t_1} \cdots \sum_{t_k} \circ$:

\vspace{-3mm}
\begin{align}
  = & \argmax_{s_k} \!\sum_{T_k} \Bigl( P(t|\vec{q}) \,P(t_k|s_k) \prod_{i=1}^{k-1} P(t_i|s_i^*) \nonumber \\[-3mm]
  & \hspace{21mm} \, \cdot P(R_k\geq n|T_k,S_{k-1}^*,s_k,\vec{q}) \Bigr) \nonumber 
\end{align}

\vspace{-1mm}
We split $R_k \geq n$ into two disjoint (additive) events
$(r_k \! \geq \! 0,\! R_{k\!-\!1}\!\geq \!n)$, $(r_k\!\!=\!\!1,\!R_{k\!-\!1}\!\!=\!\!n\!-\!1)$ where all $r_i$ are D-separated:

\vspace{-3.5mm}
\begin{align}
  = & \argmax_{s_k} \!\sum_{T_k} P(t|\vec{q}) \,P(t_k|s_k) \prod_{i=1}^{k-1} P(t_i|s_i^*) \nonumber \\
  & \hspace{6mm} \cdot \Bigl( \mbox{$\underbrace{P(r_k\!\geq\!0|R_{k-\!1}\!\geq\!n,t_k,t)}_{1}$} P(\!R_{k-\!1}\!\geq\!n|\TlessK) \nonumber \\[-1.5mm]
  & \hspace{10mm} + P(r_k=1|R_{k-1}\!=\!n\!-\!1,t_k,t) P(\!R_{k-\!1}\!=\!n\!-\!1|\TlessK) \Big) \nonumber 
\end{align}
We distribute initial terms over the summands noting that 
$\sum_{t_k} \!\! P(t_k|s_k) P(r_k\!\!=\!\!1|t_k,t) \! = \!\! \sum_{t_k} \!\! P(t_k|s_k) \I[t_k\!\!=\!\!t] \! = \!\! P(t_k\!\!=\!\!t|s_k)$:
\begin{align}
 = & \argmax_{s_k} \!\!\!\Bigg( \!\!\sum_{\TlessK} \!\!\bigg[ \mbox{$\underbrace{ \!\sum_{t_k} \!\!P(t_k|s_k) }_{1}$} \!\bigg] \!\!\!\:P(\!R_{\!k-\!1}\!\!\geq\!n|\TlessK) P(t|\vec{q}) \!\!\prod_{i=1}^{k-1} \!\!\!P(t_i|s_i^*) + \nonumber \\[-3.5mm]
  & \hspace{1mm} \sum_{t} \!P(t|\vec{q}) P(t_k\!=\!t|s_k) \hspace{-4mm} \sum_{t_1, \dots, t_{k-1}} \hspace{-4mm} P(R_{k-\!1}\!=\!n\!-\!1|\TlessK) \!\prod_{i=1}^{k-1} \!P(t_i|s_i^*) \!\!\Bigg) \nonumber
\end{align}
Next we proceed
to drop the first summand since it is not a function of $s_k$ (i.e.,
it has no influence in determining $s_k^*$):
% This give us the
%simplified optimization objective:
\begin{align}
= & \argmax_{s_k} \!\sum_{t} \!P(t|\vec{q}) P(t_k\!=\!t|s_k) P(\!R_{k-\!1}\!\!=\!n\!-\!1|S_{k-1}^*) \label{eq.ncall}
\end{align}
By similar reasoning, we can derive that the last probability 
needed in~\eqref{eq.ncall} is recursively defined as $P(R_k=n|S_k,t)=$
\begin{align*}
\begin{cases}
n \geq 1, k > 1:  &  \bigl( 1\!-\!P(t_k\!=\!t|s_k) \bigr) P(R_{k-1}\!=\!n|S_{k-1},t) \nonumber \\
  & \hspace{5mm} + P(t_k\!=\!t|s_k) P(R_{k-\!1}\!=\!n\!-\!1|S_{k-\!1},t) \\
n = 0, k > 1:   & \bigl( 1\!-\!P(t_k\!=\!t|s_k) \bigr) P(R_{k-\!1}\!=\!0|S_{k-\!1},t) \\
n = 1, k = 1:   & P(t_1\!=\!t|s_1) \\
n = 0, k = 1:   & 1 - P(t_1\!=\!t|s_1)
\end{cases}
%       \bigl 1-P(t_1\!=\!t|s_1) \bigr) = \bigl 1-P(t_1\!=\!t|s_1) \bigr) \\
%  & \hspace{2mm} P(R_1\!=\!1|S_1,t) = P(t_1\!=\!t|s_1)
\end{align*}
We can now rewrite~\eqref{eq.ncall} by unrolling its recursive definition.
For expected $n$-call@$k$ where $n \leq k/2$ %, n \! \neq \! 1$ 
(a symmetrical result holds for $k/2 < n \leq k$), the explicit unrolled objective is 
\begin{align}
  & s_k^* = \argmax_{s_k} \sum_t \Biggl( P(t|\vec{q}) \, P(t_k=t|s_k) \cdot \nonumber \\[-2.5mm]
  & \hspace{10mm} \sum_{j_1, \dots, j_{n-\!1}} \hspace{-14mm} \prod_{\hspace{14.5mm} l \in \{j_1, \dots, j_{n-\!1}\}} \hspace{-14mm} P(t_l\!=\!t|s_l^*) \hspace{-13mm} \prod_{\substack{i=1 \\ \hspace{14mm} i \notin \{j_1, \dots, j_{n-\!1}\}}}^{k-1} \hspace{-13mm} \!\bigl( 1 - P(t_i\!=\!t|s_i^*) \bigr) \!\Biggr) \label{eq.ncall.alt}
\end{align}
where $j_1, \dots, j_{n-1} \in \{1,\ldots,k-1\}$ satisfy 
that $j_i < j_{i+1}$ (i.e.,
an ordered permutation of $n-1$ result set indices).
%A similar objective can be easily obtained for the case $n>k/2$,
%$n\neq k$ \emph{via} the same process (not shown due to space).

% ===============================================================================

%From here we focus on the last product in~\eqref{eq.ncall.alt}.  We note that
If we assume each document covers a single subtopic of the query (e.g.,
a subtopic represents an intent of an ambiguous query) then we can assume that 
$\forall i \; P(t_i|s_i) \in \{0,1\}$ and $P(t|\vec{q}) \in \{0,1\}$.  This
allows us to convert a $\prod$ to a $\max$ 
\begin{align*}
  \hspace{-13mm} \prod_{\substack{i=1 \\ \hspace{14mm} i \notin \{j_1, \dots, j_{n-\!1}\}}}^{k-1} \hspace{-14mm} \bigl( 1 - \!P(t_i\!=\!t|s_i^*) \bigr) & =
1 - \Biggl( 1 - \hspace{-14mm} \prod_{\substack{i=1 \\ \hspace{14mm} i \notin \{j_1, \dots, j_{n-\!1}\}}}^{k-1} \hspace{-13mm} \bigl( 1 - P(t_i\!=\!t|s_i^*) \bigr) \Biggr) \\[-2mm]
  & = 1 - \Bigl( \hspace{-6mm} \max_{\substack{i \in [1,k-1] \\ \hspace{7mm} i \notin \{j_1, \dots, j_{n-1}\}}} \hspace{-6mm} P(t_i\!=\!t|s_i^*) \Bigr)
\end{align*}
and by substituting this into~\eqref{eq.ncall.alt} and distributing, we get
\vspace{-1.5mm}
\begin{align}
=  & \, \argmax_{s_k} \sum_t \Biggl( P(t|\vec{q}) P(t_k\!=\!t|s_k)  \sum_{\hspace{-1mm} j_1, \dots, j_{n-\!1}} \hspace{-13.5mm} \prod_{\hspace{14.5mm} l \in \{j_1, \dots, j_{n-1}\}} \hspace{-14mm} P(t_l\!=\!t|s_l^*) \nonumber \\[-2mm]
  & \hspace{-1mm} - \!P(t|\vec{q}) P(t_k\!=\!t|s_k) \hspace{-1mm} \sum_{\hspace{-5mm} j_1, \dots, j_{n-\!1}} \hspace{-14mm} \prod_{\hspace{14.5mm} l \in \{j_1, \dots, j_{n-\!1}\}} \hspace{-15mm} P(t_l\!=\!t|s_l^*) \hspace{-11.5mm} \max_{\substack{\hspace{5mm} i \in [1,k-1] \\ \hspace{11.5mm} i \notin \{j_1, \dots, j_{n-\!1}\}}} \hspace{-9.5mm} P(t_i\!=\!t|s_i^*) \!\!\Biggr) . \nonumber %\\[-2mm]
%  & \hspace{3mm} - \!P(t|\vec{q}) P(t_k\!=\!t|s_k) \hspace{-1mm} \mbox{$\underbrace{\sum_{\hspace{-5mm} j_1, \dots, j_{n-\!1}} \hspace{-14mm} \prod_{\hspace{14.5mm} l=\{j_1, \dots, j_{n-\!1}\}} \hspace{-15mm} P(t_l\!=\!t|s_l^*) \hspace{-11.5mm} \max_{\substack{\hspace{5mm} i=[1,k-1] \\ \hspace{11.5mm} i \notin \{j_1, \dots, j_{n-\!1}\}}} \hspace{-9.5mm} P(t_i\!=\!t|s_i^*)}_{}$} \!\!\Biggr) \nonumber \\[-3.5mm]
%  & \hspace{40mm} \max_{s_i \in S_{k-1}^*} P(t_i\!=\!t|s_i) \, w_i \nonumber\\
\end{align}
Assuming $m$ selected documents $S_{k-1}^*$ are relevant 
then the top term
(specifically $\prod_l$) is non-zero $\binom{m}{n-1}$ times.  For the
bottom term, it takes $n-1$ relevant $S_{k-1}^*$ to satisfy its
$\prod_l$, and one additional relevant document to satisfy the
$\max_i$ making it non-zero $\binom{m}{n}$ times.  Factoring out the
$\max$ element from the bottom and pushing the $\sum_t$ inwards (all legal
due to the $\{0,1\}$ subtopic probability assumption) we get
\vspace{-1.5mm}
\begin{align}
=  & \argmax_{s_k} \binom{m}{n-1} \underbrace{\sum_t P(t|\vec{q}) P(t_k\!=\!t|s_k)}_{\textrm{relevance}: \; \Sim_1(s_k,\vec{q})} \nonumber \\[-3mm]
  & \hspace{10.5mm} - \binom{m}{n} \max_{s_i \in S_{k-1}^*} \underbrace{\sum_t P(t_i\!=\!t|s_i) \!P(t|\vec{q}) P(t_k\!=\!t|s_k)}_{\textrm{diversity}: \; \Sim_2(s_k,s_i,\vec{q})} \nonumber .\\[-7mm] \nonumber
%=  & \argmax_{s_k} \sum_t \Bigg[ \binom{m}{n-1} P(t|\vec{q}) P(t_k\!=\!t|s_k) \nonumber \\[-2mm]
%  & \hspace{17.5mm} - \binom{m}{n} \max_{s_i \in S_{k-1}^*} P(t_i\!=\!t|s_i) \!P(t|\vec{q}) P(t_k\!=\!t|s_k) \Bigg] \nonumber
\end{align}  
From here we can 
normalize by $\binom{m}{n-1} + \binom{m}{n} = \binom{m+1}{n}$ 
(Pascal's rule), leading to fortuitous cancellations and the result:
\vspace{-3mm}
\begin{align}
=  & \argmax_{s_k} \!\! \frac{n}{m\!+\!1} \Sim_1(s_k,\vec{q}) - \frac{m\!-\!n\!+\!1}{m+1} \max_{s_i \in S_{k-1}^*} \! \Sim_2(s_k,s_i,\vec{q}) \nonumber \\[-6mm] \nonumber
\end{align}
%Fortuitously, we note that the $\binom{m+1}{n}$ divisor cancelled with
%the numerators, yielding this elegant and interpretable result.
Comparing to MMR in~\eqref{eq:MMR}, we can clearly see that $\lambda =
\frac{n}{m\!+\!1}$.  Assuming $m \approx n$
since \ExpNCall{n} optimizes for the case where $n$ relevant documents are selected, then $\lambda = \frac{n}{n\!+\!1}$. 
%, which achieves our goal of formally expressing the 
%relevance vs. diversity tradeoff as a function of $n$, $k$, and $m$.

%As a reality check, we see that this coincides
%with the published result of $\lambda=0.5$ in~\cite{sanner11} for
%$n=1$, $m=1$.  Overall we have achieved our goal and have shown that
%indeed, diversificiation in expected $n$-call@$k$ decreases linearly 
%as $n \to 1$.

% ===============================================================================

\vspace{-0.75mm}

{\small 

\section*{Acknowledgements}

NICTA is funded by the Australian Government via 
the Dept. of Broadband, Comm. and the Digital
Economy and the Australian Research Council through the ICT
Centre of Excellence program.}

\vspace{-0.75mm}

\bibliographystyle{abbrv}
%\bibliography{reference}

\end{document}